# Formation dynamics and distribution function of cities population

B. R. Gadjiev, M. A. Korolev and T. B. Progulova

International University for Nature, Society and Man,
19 Universitetskaya Street, 141980 Dubna, Russia

*From the data analysis we defined distribution function against the population on the level of various structure units, namely regions, federal districts and the country on the whole. We have studied peculiarities of the distribution function deformation due to the structure units' enlargement. Using the master equation in the continuous approximation, we obtain the Fokker-Plank equation for the distribution function with symmetric transition rates. In addition, we offer a model where transitions only between neighbour states are possible. Moreover in this case it is proposed the in and out transition probability rate for any states are different We define condition for the formal equivalence of both models to the problem that is described by the stochastic differential equation with additive and multiplicative white noises. We have analyzed the corresponding Fokker-Plank equation in the Ito and Stratonovich calculus, and obtained a solution of the Tsallis distribution type. We demonstrate a comparison of theoretical results with those of the data analysis on the level a cumulative distributions. We have used the maximum likelihood method for fitting. The obtained results allow us to attribute the specific value of the Tsallis distribution nonextensivity parameter to the empiric curves.*

**Introduction.** Population distribution in cities is investigated in a number of papers where it is shown that this distribution follows the Zipf law with an exponent $\gamma \approx 2$ [1, 2]. Apparently, this law has a universal character and reveals both on the world level and for separate countries. The determination of the exponent $\gamma$ for many countries shows that $\gamma$ weakly depends on social and economic status of the population. The $\gamma$ value for 2400 US cities is $\gamma \in (2.1 \pm 0.1 \div 2.30 \pm 0.5)$, while for 1300 municipalities in Switzerland it is $\gamma = 2.1 \pm 0.1$ [2]. It should be mentioned that the exponents were determined by the least square method which, as it is well-known, is not the best tool to fit power law distributions. It has been determined from a more detailed fitting of population distribution in cities of Brazil with the population over 30 000 that $\gamma \approx 2.24 \pm 0.34$ for the censuses of 1970 and 1980 and $\gamma \approx 2.26 \pm 0.11$ for those of 1991 and 2000. Population distribution for 2700 cities of the world with the population more than 100 000 also reveals the Zipf law with exponent $\gamma \in 2.03 \pm 0.05$ [3]. Therefore, from the point of view of the general system theory we deal with a manifold of interacting cities on various levels (separate countries and the world), which, independent of the microscopic dynamics details, reveals the power law characteristic for complex systems. Unfortunately, all the above mentioned papers determine the exponent $\gamma$ from the rectilinear region of the distribution plotted in the log-log scale. As a matter of fact, it is more correct to describe these distributions not by the Zipf law but at least by the Zipf-Mandelbrot law.

In this paper, we present empiric data on population distribution in Russian cities on the level of various administrative units, namely, the Moscow region, the Central Federal District and on the level of the whole country. We give the data analysis results on population distribution. We apply the maximum entropy approach for complex systems and obtain the distribution function with which we accomplish the empiric distributions' fitting. Hereafter, we present stochastic models of the cities population growth that describe the behaviour observed empirically.

**Data analysis.** We have analyzed data on population for 1 January 2006 published by the Federal Office of State Statistics. De jure population of the Russian Federation was in this period about 143,1 million people, with 104,3 million people (or 72.9%) living in cities and 38,8 million people (or 27.1%) being residents of non-urban areas. At that time, there were 1095 cities in Russia. In 167 cities the population was more than 100 000 people. The population of the Central Federal District was 37.4 million people. The urban population of 307 cities was 30 million people (or 80.3%). The population of the Moscow region (without the city of Moscow which is a separate administrative unit of the federal subordination) in the same period of time was about 6.628 million people, including 5.328 million people (80.4%) being urban citizens. There are 80 cities in the Moscow region; many of them have population over 100 000 people. There are two types of settlements in the Moscow region – summer communities and settlements of urban type; the latter have larger population. Population of some settlements is more than 20 000 people and due to this fact they are traditionally treated as city population as well. Therefore, we used data not only on cities but on all inhabited localities to enlarge statistics on the Moscow region.

We have constructed the population distributions of cities on three levels: for the Russian Federation as a whole, for the Central Federal District and the Moscow region. The results are presented in Fig. 1 in the log-log scale. The first characteristic feature is an obvious and considerable deviation of the obtained distributions from the Zipf law, especially on the levels of the Moscow region and the Central Federal District. The second peculiarity is the apparent presence of the finite-dimensional effect that reveals in the form of distribution tails. Finally, the third feature consists that with transferring from a regional level to federal one, obviously, there is a formation of a rectilinear area of distribution. Statistics restrictions in distribution tails lead to a necessity to apply the maximum likelihood method for the distribution fitting. Limitation of statistics in tails of distributions prompts necessity of use for fitting the maximum likelihood method. As we deal with a complex system, it is necessary to apply the maximum nonadditive entropy approach to define the type of the distribution function.

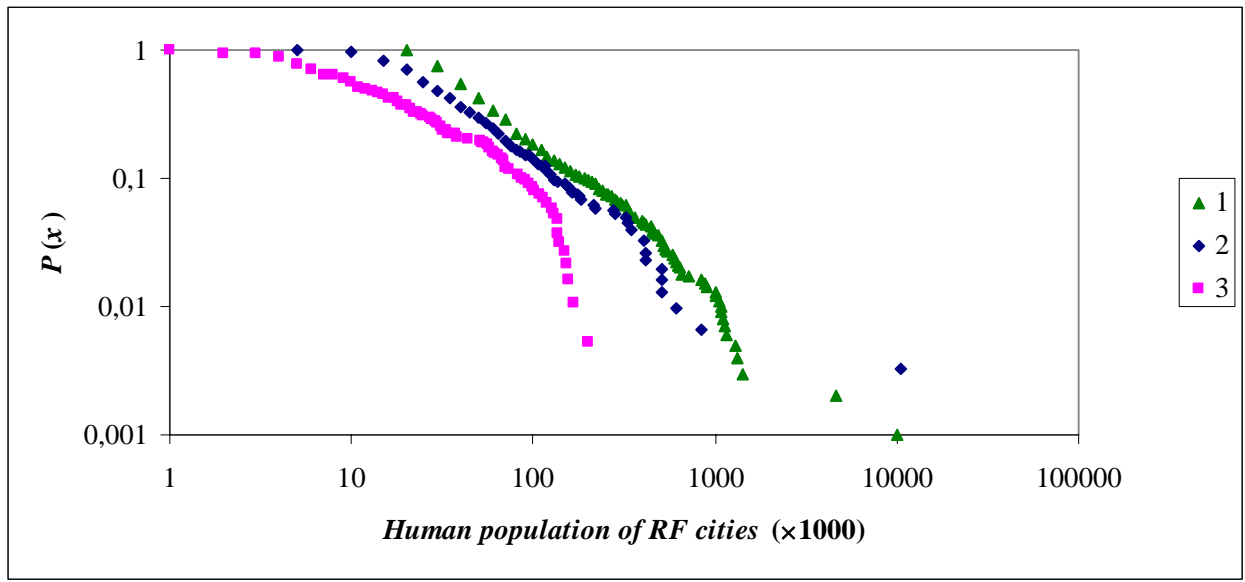

Fig. 1. Population distribution on the log-log scale for RF — 1, for the Central Federal District – 2, for the Moscow region – 3; the *q* parameter values for 1 — $q = 1.49$; for 2 — $q = 1.48$; for 3 — $q = 1.43$

**Maximum entropy approach.** To define the distribution functions of complex systems we can use the maximum-entropy method, proposed for the first time by Tsallis:

$$S_q = \frac{\sum_k p_k^q - 1}{1 - q} \qquad (1)$$

that in the $q \to 1$ limit leads to the Shannon entropy [4]. Here $p_k$ stands for the probability distribution function. The distribution function is defined from maximum equation (1) with

additional constraints. The first constraint relates to the usual condition of probabilities normalization

$$\sum_k p_k = 1. \quad (2)$$

The additional constraint laid on the nonadditive entropy, has the form

$$\sum_k P(k)k = \mu \quad (3)$$

with

$$P(k) = \frac{p_k^q}{c_q} \quad (4)$$

where $P(k)$ is the escort probability and

$$c_q = \sum_k p_k^q \quad (5)$$

Taking these constraints into account to define the entropy maximum, we can use the Lagrange multipliers method

$$\frac{\delta}{\delta p_k}\left[-\frac{\sum_{k'} p_{k'}^q - 1}{1-q} + \lambda \sum_{k'} p_{k'} + \varsigma \sum_{k'} k' P(k')\right] = 0. \quad (6)$$

It is easy to show that

$$p_k = \left[(1-q)\frac{\lambda}{qk_0\varsigma}\right]^{\frac{1}{q-1}}\left(1-(1-q)\frac{k}{k_0}\right)^{\frac{1}{1-q}}, \quad (7)$$

where

$$k_0 = \frac{1}{\varsigma}c_q\left(1 + \varsigma(1-q)\frac{\mu}{c_q}\right) \quad (8)$$

The $\lambda$ parameter is defined from normalization condition (2), namely

$$\left(\frac{\lambda}{q\varsigma k_0}(1-q)\right)^{\frac{1}{q-1}} = \left[\sum_k\left(1-(1-q)\frac{k}{k_0}\right)^{\frac{1}{1-q}}\right]^{-1} \quad (9)$$

Introducing the notations

$$Z = \sum_k\left(1-(1-q)\frac{k}{k_0}\right)^{\frac{1}{1-q}} \quad (10)$$

for the probabilities distribution $p_k$ we represent it in the form

$$p_k = \frac{1}{Z}\left(1-(1-q)\frac{k}{k_0}\right)^{\frac{1}{1-q}} \quad (11)$$

We used the Tsallis distribution (11) for fitting the data processing results that are given in Fig.1.

**Stochastic differential equation approach I.** Assume that a social system consists of population divided into groups $i = 1,2,...$, with a arbitrary number of groups. Such division may be determined by living in locality $i$. We specify as $n_i$ the quantity of persons living in locality $i$. Then the population configuration takes the form $\vec{n} = \{...,n_i,...\}$. Assume that transitions between states are possible $\{...,n_i,...\} \Rightarrow \{...,n_i+1,...\}$ and $\{...,n_i,...\} \Leftarrow \{...,n_i+1,...\}$ and $\{...,n_i,...\} \Rightarrow \{...,n_i-1,...\}$ and $\{...,n_i,...\} \Leftarrow \{...,n_i-1,...\}$. Then the master equation has the form:

$$\frac{dp(n,t)}{dt} = a(n+1)p(n+1,t) + b(n-1)p(n-1,t) - a(n)p(n,t) - b(n)p(n,t), \quad (12)$$

where $a(x)$ and $b(x)$ are the transitions probability rates between corresponding states. Inherently, the operator translation is $f(x+\alpha) = e^{\alpha D} f(x)$, where $D = \dfrac{\partial}{\partial x}$ [5]. In the continuous limit, using the operator translation equation (12) may be written in the form,

$$\frac{dp(n,t)}{dt} = e^{D}(a(n)p(n,t)) + e^{-D}(b(n)p(n,t)) - a(n)p(n,t) - b(n)p(n,t) \tag{13}$$

Using expansion $e^{D} = 1 + \dfrac{1}{1!}\dfrac{\partial}{\partial x} + \dfrac{1}{2!}\dfrac{\partial^2}{\partial x^2} + ...$, equation (12) can be represented in the form

$$\frac{\partial p(n,t)}{\partial t} = \frac{\partial}{\partial n}[(a(n)-b(n))p(n,t)] + \frac{1}{2}\frac{\partial^2}{\partial n^2}[(a(n)+b(n))p(n,t)] \tag{14}$$

Introducing the notations $D_1(x) = b(x) - a(x)$ and $D_2(x) = \dfrac{1}{2}(b(x)+a(x))$, from equation (14) we obtain

$$\frac{\partial p(n,t)}{\partial t} = -\frac{\partial}{\partial n}[D_1(n)p(n,t)] + \frac{\partial^2}{\partial n^2}[D_2(n)p(n,t)] \tag{15}$$

Assume that

$$D_1(n) = -\frac{1}{2}\tau + \frac{1}{2}M, \quad D_2(n) = Mn + A \tag{16}$$

The probability current is

$$j = -\left(\frac{1}{2}\tau p(n,t) - \frac{1}{2}Mp(n,t) - \frac{\partial[(A+Mn)p(n,t)]}{\partial n}\right) \tag{17}$$

and then equation (15) can be written in the form

$$\frac{\partial p(n,t)}{\partial t} = -\frac{\partial}{\partial n}j \tag{18}$$

Taking the boundary conditions into account [5,6]
$$p(n=\infty) = p(n=-\infty) = 0$$
we obtain
$$j(n=\infty) = j(n=-\infty) = 0.$$

Let us define the stationary solution for equation (4), namely, we assume that $\dfrac{\partial p(n,t)}{\partial t} = 0$ and, consequently, we obtain

$$(\tau + M)p_{st}(n) + (A+Mn)\frac{\partial p_{st}(n)}{\partial n} = 0 \tag{19}$$

Equation (8) admits the variables separation

$$\frac{dp_{st}(n)}{p_{st}(n)} = -\frac{\tau+M}{2}\frac{dn}{(A+Mn)} \tag{20}$$

Equation (10) integrating leads to

$$p_{st}(n) = p_0(A+Mn)^{-\frac{\tau+M}{2M}}, \tag{21}$$

where $p_0$ is the integrating constant.
Introducing the notations,

$$q = \frac{3M+\tau}{M+\tau}, \quad \frac{M}{A} = (q-1)\beta, \tag{22}$$

we can write equation (21) in the form

$$p_{st}(n) = \frac{1}{Z}(1-(1-q)\beta n)^{\frac{1}{1-q}}, \tag{23}$$

where

$$Z = \int_0^\infty (1-(1-q)\beta n)^{\frac{1}{1-q}} dn \qquad (24)$$

In this case, in accordance with equation (14)

$$\frac{1}{2}(a(x)+b(x)) = Mx + A \quad \text{and} \quad a(x)-b(x) = \frac{1}{2}(\tau - M).$$

Thus, we come to the result obtained from the maximum entropy approach.

Stochastic differential equation approach II. The master equation for probability $P(n,t)$ of the system in the state characterized by the variable $n$ at the time moment $t$, has the form [5,6]

$$\frac{\partial P(n,t)}{\partial t} = \sum_{m=-\infty}^{\infty} [P(m,t)w(m,n;t) - P(n,t)w(n,m;t)] \qquad (25)$$

where $w(k,l;t)$ is a transition rate from the $k$ state to the $l$ state in the time interval $(t, t+\Delta t)$. For the continuous stochastic variable $x$ equation (25) takes the form

$$\frac{\partial P(n,t)}{\partial t} = \int_{-\infty}^{\infty} P(x',t)w(x'|x)dx' - P(x,t)\int_{-\infty}^{\infty} w(x|x')dx', \qquad (26)$$

where $w(y|z)$ is the intensity of transition rates from the $y$ state into the $z$ state.

Let us define the transition rates between the states $x$ and $x+y$ by $\phi(x,y) = w(x|x+y)$. Then equation (26) can be written in the form

$$\frac{\partial P(n,t)}{\partial t} = \int_{-\infty}^{\infty} P(n-y,t)\phi(n-y,y)dy - P(n,t)\int_{-\infty}^{\infty} \phi(n,y)dy \qquad (27)$$

Let us assume that $\phi(x,y)$ is a symmetric function. Using the definition of the operator translation

$$P(n-y,t)\phi(n-y,y) = e^{-yD}[P(n,t)\phi(n,y)] = P(n,t)\phi(n,y) + \sum_{l=1}^{\infty} \frac{(-y)^l}{l!} \frac{\partial^l}{\partial n^l}(P(n,t)\phi(n,y)) \qquad (28)$$

and introducing the notation

$$\chi_l(n) = \int_{-\infty}^{\infty} y^l \phi(n,y) dy \qquad (29)$$

we present equation (27) in the form

$$\frac{\partial P(n,t)}{\partial t} = \sum_{l=1}^{\infty} \frac{(-1)^l}{l!} \frac{\partial^l}{\partial n^l}(P(n,t)\chi_l(n)) \qquad (30)$$

Assume that all $\chi_l$ with $l \geq 3$ are equal to zero. Then equation (30) takes the form

$$\frac{\partial P(n,t)}{\partial t} = -\frac{\partial}{\partial n}(P(n,t)\chi_1(n)) + \frac{1}{2}\frac{\partial^2}{\partial n^2}(P(n,t)\chi_2(n)), \qquad (31)$$

where, according to (29)

$$\chi_1(n) = \int_{-\infty}^{\infty} y\phi(n,y)dy \qquad (32)$$

and

$$\chi_2(n) = \int_{-\infty}^{\infty} y^2 \phi(n,y)dy \qquad (33)$$

Assume that

$$\chi_1(n) = -\frac{1}{2}\tau + \frac{1}{2}M \qquad (34)$$

and

$$\chi_2(n) = 2(A + Mn) \tag{35}$$

In this case equation (31) formally is similar to equation (15), and, consequently, its solution is represented by expression (23).

**Multiplicative noise: A mechanism leading to nonextensive statistical mechanics.** Let us consider the Langevin equation in the form [4]

$$\frac{dx}{dt} = -\frac{\tau}{2} + \sqrt{x}\xi(t) + \eta(t), \tag{36}$$

where $x(t)$ is a stochastic variable and $\xi(t)$ and $\eta(t)$ are noncorrelated noises with mean value equal to zero

$$\langle \xi(t)\xi(t')\rangle = 2M\delta(t-t'), \tag{37}$$

$$\langle \eta(t)\eta(t')\rangle = 2A\delta(t-t'), \tag{38}$$

where $M > 0$ and $A > 0$ are amplitudes of the multiplicative and additive noises, respectively.

The Fokker-Plank equation for probability density, connected to equation (36) in the Stratonovich calculus has the form

$$\frac{\partial p(x,t)}{\partial t} = \frac{\tau}{2}\frac{\partial (p(x,t))}{\partial x} + \frac{M}{2}\frac{\partial}{\partial x}\left(\sqrt{x}\frac{\partial}{\partial x}\left(\sqrt{x}p(x,t)\right)\right) + A\frac{\partial^2 p(x,t)}{\partial x^2} \tag{39}$$

It is easy to show that the stationary solution of this equation takes the form

$$p_{st}(x) = p_0 A^{-\frac{\tau+M}{2M}}\left(1 + \frac{M}{A}x\right)^{-\frac{\tau+M}{2M}}, \tag{40}$$

and, consequently,

$$p_{st}(x) = \frac{1}{Z}(1-(1-q)x)^{\frac{1}{1-q}} \tag{41}$$

Thus, the multiplicative noise effect leads to the nonextensivity system. The formal mathematical equivalence of the above discussed models with multiplicative noise shows that these models describe nonextensive systems as well.

We should mark that as inherently $q = \frac{3M + \tau}{M + \tau}$, at $M = 0$ we obtain $q = 1$. In this case $p_{st}(x)$ results in exponential distribution, and, consequently, it is possible to state that the system nonextensivity is connected to the effect of multiplicative noises on the system. Besides, the obtained distribution at large $x$ results in the power law distribution. If $x \leq x_{max}$ and $q \neq 1$, behaviour occurs that is intermediate between the exponential and power law one.

**Conclusion.** Systems with the additive entropy ($S_1 = -k\int du p(u)\ln p(u)$) are ergodic in all phase space of the variables and are generally described with exponential distributions. In complex systems that are characterized by the nonadditive entropy some subspaces in phase space turn out to be more *preferable* than others, and an account of such subspaces is essential. It is suitable to describe such systems in the frames of nonextensive statistical mechanics which is based on the nonadditive entropy $S_q = k\left[\dfrac{1 - \int du [p(u)]^q}{q-1}\right]$. A wide family of models described by the multiplicative noise belongs to the nonextensivity type. The system variables are directly connected to noise in the presence of the multiplicative noise. Thus, we observe in such systems the behaviour that does not occur in the presence of only additive noise. In particular, the interaction between the additive and multiplicative noises may lead to the emergence of the $q$-exponent.

In the system of interacting cities the social-geographical conditions of the country lead to the situation when some cities are more preferable for living than others. Consequently, the population distribution of the cities is described by the Tsallis distribution which amounts at large values of the urban population to the Zipf law. This effect does not reveal considerably when regarding relatively small regions (curve 1 in Fig. 1). In this case, the population distribution is described with the $q$-exponential function with restricted argument, so that the power law region in the log-log dependence is absent. It is related to the restrictions on the cities spaciousness. Partial extension of the inhabited region deforms the population distribution function. (curve 2 in Fig. 2). Further deformation of the distribution function occurs with further inhabitance space extension, and a rectilinear region appears in the log-log scale being a feature of the power law dependence. We have defined the exponent value of the Zipf law $\gamma = 2.04$ from the rectilinear part of curve 1, Fig. 1 by the maximum likelihood methods fitting. Upon that, all dependence is described by the Tsallis distribution and a corresponding values of entropic indexes are 1 — $q = 1.49$; for 2 — $q = 1.48$; for 3 — $q = 1.43$.

**Reference**

1. Zanette D. H. and Manrubia S. C. Role of intermittency in urban development: A model of large-scale city formation, Phys. Rev. Lett. 79 (1997) 523-526.
2. Newman M. E. J., Power laws, Pareto distributions and Zipf's law, arXiv:cond-mat/0412004.
3. Newton J., Moura Jr., Marcelo B. Ribeiro Zipf law for Brazilan cities, Physica A 367 (2006) 441-448.
4. Anteneodo C. and Tsallis C. Multiplicative noise: A mechanism leading to nonextensive statistical mechanics, Journal of Mathematical Physics 44 (2005) 5194-5203.
5. Reichl L. E., *A modern course in statistical physics*, 2nd ed. (Wiley, New York, 1998)
6. Evaldo M., Curado F., Nobre F. D. Derivation of nonlinear Fokker-Plank equations by means of approximations to the master equation Phys. Rev. E 67 (2003) 021107-1 – 021107-7.